\documentclass[%
 reprint,
 superscriptaddress,
 showpacs,preprintnumbers,
 amsmath,amssymb,
 aps,
 prl,
 longbibliography,
lengthcheck,%
]{revtex4-1}

\usepackage{graphicx}
\usepackage{dcolumn}
\usepackage{bm}
\usepackage{color}
\usepackage{ulem}
\usepackage{hyperref}
\usepackage{braket}

\usepackage{natbib}

\makeatletter
\def\@cite#1#2{（#1\if@tempswa , #2\fi）}
\makeatother
\DeclareMathOperator{\trace}{Tr}
\renewcommand{\Im}{\mathrm{Im}~}
\renewcommand{\Re}{\mathrm{Re}~}
\newcommand{\dd}[1]{\mathrm{d}#1}

\newcommand{\pdv}[2]{\frac{\partial #1}{\partial #2}}

\renewcommand{\epsilon}{\varepsilon}

\begin{document}

\preprint{APS/123-QED}

\title{
Kondo Effect in Nonreciprocal Response 
}

\author{Hajime Murata}
\affiliation{
Department of Physics, Institute of Science Tokyo, Meguro, Tokyo, 152-8551, Japan
}

\author{Hiroaki Ishizuka}
\affiliation{
Department of Physics, Institute of Science Tokyo, Meguro, Tokyo, 152-8551, Japan
}

\date{\today}

\begin{abstract}
Chiral magnetic states give rise to rich phenomena, from the anomalous Hall effect and the nonlinear electrical current to multiferroics and magnetochiral dichroism.
Most of the studies on electrical transport so far have focused on the cases where the magnetic moments are well approximated by classical local moments.
Here, we reveal that the coexistence of quantum fluctuations and chiral spin correlations gives rise to a $\log(T)$ temperature dependence in the electrical magnetochiral effect, a nonreciprocal response.
Using the Green's function method and a scattering theory approach, we show that the $\log(T)$ temperature dependence occurs through a scattering process similar to that of the Kondo effect.
The electrical magnetochiral effect is sensitive to the sign of vector spin chirality and the magnetic field.
The results demonstrate that local spin correlations and quantum fluctuations cooperatively induce nontrivial properties in transport phenomena.
\end{abstract}

\pacs{
}

\maketitle


%
%

\begin{figure}
  \includegraphics[width=0.8\linewidth]{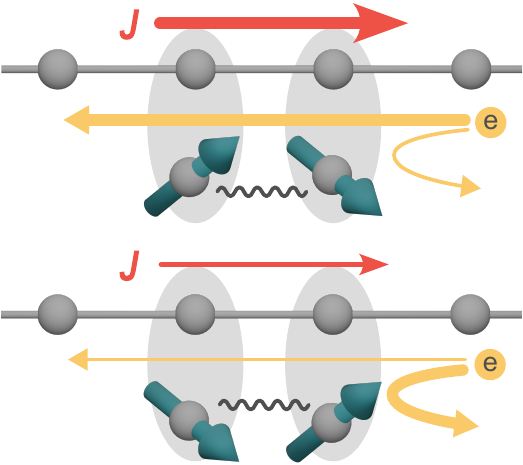}
  \caption{
    Schematics of the asymmetric magnetic scattering by two spins with non-zero vector spin chirality.
    The scattering rate for backscattering depends on the direction of the incoming electron and the vector spin chirality, resulting in a nonreciprocal current.
  }
  \label{fig:kondo_emche}
\end{figure}

{\it Introduction.---}Non-collinear magnetic states in metals lead to a variety of fascinating material properties and functionalities.
These are often related to the scalar ($\chi_{ijk}=\bm S_i\cdot\bm S_j\times\bm S_k$) and vector ($\bm \chi_{ij}=\bm S_i\times\bm S_j$) spin chiralities, where $\bm S_i$ is the magnetic moment at the $i$-th site.
For example, scalar spin chirality is known to induce a novel anomalous Hall effect (AHE)~\cite{Ye1999a,Ohgushi2000a,Tatara2002a}.
In the presence of spin-orbit interaction, a distinct contribution to the AHE also emerges from the vector spin chirality due to the lower spin rotation symmetry~\cite{Ishizuka2018b, Zhang2018a, Lux2020a, Kipp2021a,Yamaguchi2021a, Mochida2024a, Terasawa2024a}.
The chiral spin state also induces nonreciprocal responses~\cite{Tokura2018a}, often characterized by an electric current proportional to the square of the applied electric field $J\propto E^2$~\cite{Rikken2001a,Rikken2005a}. 
For example, the electrical magnetochiral effect in chiral magnets~\cite{Yokouchi2017a,Aoki2019a,Kitaori2021a} is related to the vector spin chirality~\cite{Ishizuka2020a} (Fig.~\ref{fig:kondo_emche}).
Thermally fluctuating spins also contribute to the chirality-related transport phenomena through a skew-scattering mechanism~\cite{Ishizuka2018a,Ishizuka2021a}, leading to nontrivial temperature and magnetic field dependencies~\cite{Terasawa2025a} and significantly enhanced responses~\cite{Ishizuka2021a,Fujishiro2021a}.
On the other hand, while a recent experiment has suggested the possible role of quantum fluctuations in MnSi~\cite{Yokouchi2017a,Ritz2013a}, a theoretical understanding of the quantum effects remains elusive.

%
%
The quantum nature of spins is known to induce nontrivial effects in magnets, such as the Kondo effect~\cite{Kondo1964a,Wilson1975a,Hewson1993a}.
In 1964, Kondo demonstrated that the quantum fluctuations of spins lead to temperature-dependent corrections to electron scattering, resulting in a resistivity minimum and logarithmic enhancement at low temperatures~\cite{Kondo1964a}.
Theoretical studies have also explored electron scattering involving multiple impurity spins, highlighting the potential for a non-Fermi liquid state in the presence of two impurity spins~\cite{Jones1988a,Cox1996a,Jayaprakash1981a,Aguado2000a,Jeong2001a}.
However, the interplay between chiral spin correlations and quantum fluctuations, and their role in transport phenomena, remains to be understood.

\begin{figure}
  \includegraphics[width=\linewidth]{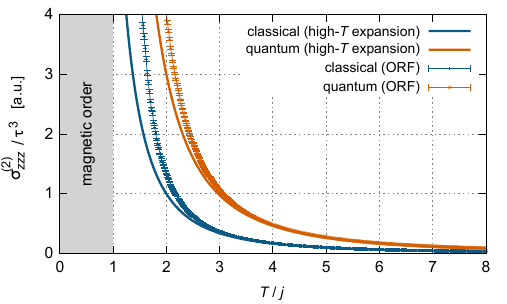}
  \caption{
    Temperature dependence of nonreciprocal conductivity $\sigma^{(2)}_{zzz}$.
    The blue lines represent the leading-order contribution, which is proportional to the product of magnetization and vector spin chirality, $m\chi^z$.
    The orange lines are ones with quantum Kondo-type corrections $2J\rho\log(T/\Lambda)$.
    The symbols connected by lines are calculated by Onsager's reaction field theory (ORF) with classical spins, while the bold solid lines are calculated by the high-temperature expansion with quantum spins.
    The results are for $d/j=0.2$, $h/j=0.002$, $T_K/j=2$, $J\rho=-0.15$, $S=1/2$ where $T_K=\Lambda\exp(1/J\rho)$ is the Kondo temperature.
  }
  \label{fig:kondo_Tdep}
\end{figure}

%
%
In this work, we study the quantum effect on the nonreciprocal response in a magnet with chiral spin correlations.
Using two different methods, the Green's function method and a Boltzmann theory combined with scattering theory, we show that the quantum correction gives rise to a characteristic logarithmic temperature dependence of the nonlinear conductivity $\sigma^{(2)}$.
The microscopic mechanism is similar to that of the Kondo effect~\cite{Kondo1964a}, in which a scattering process in the third order of the Kondo coupling gives rise to the $\log(T)$ behavior (Fig.~\ref{fig:kondo_Tdep}).
The results demonstrate a nontrivial role of quantum fluctuations in chirality-related phenomena and establish their relation to the Kondo effect through a controlled approximation.

{\it Kondo lattice Model.---}We consider a Kondo lattice model with multiple localized moments, 
whose Hamiltonian reads
\begin{align}
  H =& H_0+H_K,  \label{eq:total-hamiltonian}\\
  H_0 =& \sum_{\bm{k},\sigma}\xi_{\bm{k}\sigma}\psi_{\bm{k}\sigma}^\dagger \psi_{\bm{k}\sigma},  \label{eq:free-hamiltonian}\\
  H_K =& -\frac{J}{N}\sum_{i,\bm{k},\bm{k}^\prime,\sigma,\sigma^\prime} \bm{S}_i \cdot \psi_{\bm{k}\sigma}^\dagger \frac{\bm{\sigma}_{\sigma\sigma^\prime}}{2}\psi_{\bm{k}^\prime\sigma^\prime} e^{i (\bm{k}-\bm{k}^\prime)\cdot\bm{R}_i}, \label{eq:kondo-exchange-interaction}
\end{align}
where $H_0$ is the free electron Hamiltonian and $H_K$ is the exchange interaction between localized moments and itinerant electrons.
Here, $\psi_{\bm{k}\sigma} (\psi_{\bm{k}\sigma}^\dagger)$ is the annihilation (creation) operator for an electron with wave number $\bm{k}$ and spin $\sigma$, $\xi_{\bm{k}\sigma}= \bm{k}^2/2m-\mu$ is its eigenenergy, $\bm{S}_i$ is the spin operator for the $i$-th localized moment satisfying the commutation relation $[S_i^\mu,S_j^\nu]=i\delta_{ij}\epsilon_{\mu\nu\lambda}S_i^\lambda$, $\bm{R}_i$ is the position of $\bm S_i$, $J$ is the Kondo coupling,
and $\bm{\sigma}=(\sigma^x,\sigma^y,\sigma^z)$ is the vector of Pauli matrices.
We set $\hbar=k_B=1$ throughout this paper.

{\it Nonreciprocal current.---}To study the nonreciprocal response of this model, we evaluate the second-order dc conductivity, $\sigma^{(2)}_{\mu\nu\lambda}$, defined by $j_{\mu}^{(2)} = \sum_{\nu\lambda}\sigma^{(2)}_{\mu\nu\lambda}E_\nu E_\lambda$ using the nonequilibrium Green's function method.
Here, $E_\nu$ is the electric field and $j_{\mu}^{(2)}$ is the electric current proportional to the square of $E_\nu$.
In the nonequilibrium Green's function method, $\sigma^{(2)}_{\mu\nu\lambda}$ reads~\cite{Joao2020a,Michishita2021a},
\begin{align}
    \sigma_{\mu\nu\lambda}^{(2)}
    &=
    \Im \frac{1}{\cal{V}}\int\frac{\dd{\omega}}{2\pi}\left[
        -f(\omega) \trace \left[j_\mu\pdv{G^R}{\omega}j_{\nu\lambda}\pdv{G^R}{\omega}\right]\right.\nonumber\\
        &\quad\left.-\pdv{f(\omega)}{\omega} \trace \left[j_\mu \pdv{G^R}{\omega}j_{\nu\lambda}(G^R-G^A)\right]
    \right]\nonumber\\
    &+\Im \frac{1}{\cal{V}}\int\frac{\dd{\omega}}{2\pi}\left[
        -f(\omega) \trace \left[j_\mu\pdv{G^R}{\omega}j_\nu G^R j_\lambda \pdv{G^R}{\omega}\right]\right.\nonumber\\
        &\quad -\pdv{f(\omega)}{\omega} \trace \left[j_\mu \pdv{G^R}{\omega}j_\nu G^R j_\lambda (G^R-G^A)\right]\nonumber\\
        &\qquad\left. + (\nu\leftrightarrow\lambda)\right],\label{eq:non-reciprocal-conductivity}
\end{align}
where $f(\omega)=1/(e^{\beta\omega}+1)$ is the Fermi distribution, $\mathcal{V}$ is the volume, $\trace(\cdots)$ denotes the summation over the momentum and spins, $G^R (G^A)$ is the retarded (advanced) Green's function,
$G^{R(A)}(\bm{k},\omega) = \left[\omega-\xi_{\bm{k}}-\Sigma^{R(A)}(\bm{k},\omega)\right]^{-1}$,
with $\Sigma^{R(A)}(\bm{k},\omega)$ being the retarded (advanced) self energy,
$j_{\mu_1\cdots\mu_n}=(-e)^n \frac{\partial^n \xi_{\bm{k}}}{\partial k^{\mu_1}\cdots\partial k^{\mu_n}}$ is the current operator, and $e>0$ is the elementary charge.
The effect of Kondo coupling is included in the self energy.
In particular, the real part of the self energy reflects the band deformation due to the Kondo coupling, $\xi_{\bm{k}} \to \xi_{\bm{k}} + \Re\Sigma^R(\bm{k},\omega)$, while the inverse of its imaginary part determines the quasiparticle lifetime $\tau_{\bm{k}}^{-1} = -2\Im\Sigma^R(\bm{k},\omega)$.

To see how the self energy affects the result, let us separate the self energy into the symmetric ($\Sigma^{R(A)}_{+}$) and asymmetric ($\Sigma^{R(A)}_{-}$) parts,
$\Sigma^{R(A)}_{\pm}(\bm{k},\omega)=\frac12\left[\Sigma^{R(A)}(\bm{k},\omega)\pm\Sigma^{R(A)}(-\bm{k},\omega)\right].$
The real part of the symmetric component is included in the band structure, and hence, not explicitly considered in the calculation.
On the other hand, the imaginary part leads to conventional symmetric scattering, often approximated by the relaxation time $\tau$.
The asymmetric part $\Sigma^{R(A)}_-(\bm{k},\omega)$ plays a key role in the nonreciprocal response.
In fact, from the form of $G^{R(A)}(\bm k,\omega)$, it is evident that if the self energy $\Sigma^{R(A)}(\bm{k},\omega)$ is an even function of the wavevector $\bm{k}$, then the summation over $\bm{k}$ yields $\sigma^{(2)}_{\mu\mu\mu} = 0$, since the integrand becomes an odd function of $\bm{k}$;
the current operator $j_\mu$ is odd, while both $j_{\mu\mu}$ and $G^{R(A)}$ are even.
Hence, we will focus on $\Sigma^{R(A)}_-$ and treat $\Sigma^{R(A)}_+$ phenomenologically.

Within the asymmetric part of the self energy, the real part corresponds to the asymmetric band deformation in $\bm k$, whereas the imaginary part corresponds to asymmetric scattering.
The effect of asymmetry in the band has been discussed in the literature, which gives rise to a nonlinear response proportional to $\tau^2$~\cite{Ideue2017a};
it appears not to exhibit effects unique to the quantum nature~\cite{suppl}.
In addition to the band-deformation effect, a recent study pointed out that nonreciprocal current arises from asymmetric magnetic scattering~\cite{Ishizuka2020a}.
Hence, in this work, we focus on the asymmetric scattering contribution, $\text{Im}\Sigma^{R(A)}_{-}$.

Up to the $J^3$ order, the nonlinear conductivity by fluctuating spins reads
\begin{align}
    &\sigma^{(2)}_{z z z}\nonumber
    =\\
    &\frac{\tau a^2\sigma^2}{e}\left(-\frac{J}{2\Lambda}\right)^2\sum_{\bm{R}} \frac{\bm{M}\cdot\bm{\chi}_{\bm{R}}}{\Lambda}\left[f_1 + \left(-\frac{J}{2\Lambda}\right) f_2\right],\label{eq:sigmazzz-green}
\end{align}
where $\sigma=\frac{e^2\tau k_F^3}{3\pi^2m}$ is the linear conductivity, $a$ is the lattice constant, $\Lambda$ is the cutoff energy, $\bm{M}$ is the Zeeman field, and $\bm{\chi}_{\bm{R}}=\langle\bm{S}(\bm{R})\times\bm{S}(\bm{0})\rangle$ is the vector spin chirality~\cite{suppl}.
The functions $f_1(k_F\bm R)$ and $f_2(k_F\bm R)$ are defined in Supplemental Material~\cite{suppl}.
$f_1$ is a temperature-independent function similar to that in the classical spin case~\cite{Ishizuka2020a}, while $f_2$ arises from purely quantum effects.

\begin{figure}
    \centering
    \includegraphics[width=\linewidth]{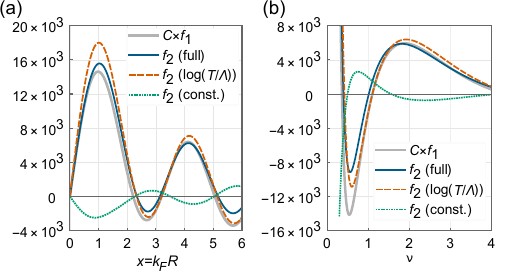}
    \caption{
        Oscillatory behavior of the dimensionless functions $f_1$ and $f_2$.
        (a) The functions plotted against the dimensionless distance $x=k_F R$.
        The solid blue line shows the full function $f_2$, which is composed of a logarithmic part (dashed orange line) and a temperature-independent part (dotted green line).
        The solid gray line shows the function $f_1$ scaled by a factor $C=6(\Lambda/\mu)[-\log(T/\Lambda)]$ for comparison with the logarithmic part of $f_2$, highlighting their similar oscillatory forms.
        (b) The same functions plotted with respect to the band filling $\nu$.
        Line styles and colors are the same as in panel (a).
        Parameters used for the plots are $\Lambda=4$ and $T=0.01$, with a chemical potential given by $\mu = 2^{1/3}\nu^{2/3}$.
        Panel (a) is plotted at a fixed filling of $\nu=2$, while panel (b) uses a fixed nearest-neighbor distance $R=a$.
    }
    \label{fig:f1f2_vs_nu}
\end{figure}

Both $f_1$ and $f_2$ are oscillating functions of $k_F\bm{R}$ as in Fig.~\ref{fig:f1f2_vs_nu}(a).
This dependence on $k_F$ implies that the sign of nonreciprocal conductivity is sensitive to material parameters such as the electron density. 
We define an effective band filling as $\nu = \frac{1}{3\pi^2}(k_F a)^3$.
Consequently, both functions also oscillate with this effective filling $\nu$ (Fig.~\ref{fig:f1f2_vs_nu}(b)).
The two functions share nearly identical nodes and overall shape, suggesting a connection between their functional forms.

A detailed analysis reveals that the temperature dependence of $f_2$ is logarithmic and that the coefficient of this logarithmic term is approximately proportional to $f_1$ (see Supplemental Material).
This logarithmic contribution is the dominant part of $f_2$.
Based on these findings, we use the approximation $f_2 \approx 6(\Lambda/\mu)\nu [-\log(T/\Lambda)] f_1$, which recasts Eq.~\eqref{eq:sigmazzz-green} into the form:
\begin{align}
    &\sigma^{(2)}_{zzz}\approx\nonumber\\
    &\frac{\tau a^2\sigma^2}{e} \left(-\frac{J}{2\Lambda}\right)^2 \left[1+2J\rho\log\left(\frac{T}{\Lambda}\right)\right] \sum_{\bm{R}} \frac{\bm{M}\cdot\bm{\chi}_{\bm{R}}}{\Lambda}f_1,\label{eq:sigma-kondo}
\end{align}
where $\rho=3\nu/2\mu$ is the density of states at the Fermi energy.
The prefactor $1+2J\rho\log(T/\Lambda)$ is the 
correction characteristic of the Kondo effect, implying that the $\log(T)$ dependence of the nonlinear response is also related to the Kondo effect.
It also implies that the renormalization of higher-order terms with respect to $J\rho\to J\rho(1+J\rho\log(T/\Lambda))$, which is known for the single-spin case~\cite{Anderson1970a,Coleman2015a}, is also valid for multiple spins.

The purely quantum origin of $f_2$ is evident from an operator-counting perspective.
While this third-order term in $J$ involves three scattering events, it is proportional to a two-spin correlator, the vector spin chirality $\bm{\chi}_{ij}=\langle\bm{S}_i\times\bm{S}_j\rangle$.
This indicates that the derivation necessarily relies on the quantum spin commutation relation $\bm{S}\times\bm{S}=i\bm{S}$, a feature with no classical analogue.

{\it Kondo Effect in Asymmetric Scattering.---}To obtain a clear physical picture of the mechanism behind Eqs.~\eqref{eq:sigmazzz-green} and \eqref{eq:sigma-kondo}, and to compare our result with the classical one~\cite{Ishizuka2020a}, we next examine the scattering probability.
To this end, we consider the scattering of an electron from a many-body state $\psi^\dagger_{\bm{k}\alpha}\ket{\text{FS}}$ to $\psi^\dagger_{\bm{k}^\prime\beta}\ket{\text{FS}}$, $W_{\bm{k}\alpha\to\bm{k}^\prime\beta}$, where $\ket{\text{FS}}$ is the Fermi sea of the non-interacting system $H_0$.
Within the second-order Born approximation, the asymmetric part contributing to the nonreciprocal current reads~\cite{Ishizuka2020a,Isobe2020a},
\begin{align}
    W^-_{\bm{k}\alpha\to\bm{k}^\prime\beta}
    \simeq&
    \sum_{i\neq j}2\pi c(\xi_{\bm{k}\alpha})i\sigma^y_{\alpha\beta}\nonumber\\
    &\times\sin(\Delta\bm{k}\cdot\bm{R}_{i j})\delta(\xi_{\bm{k}\alpha}-\xi_{\bm{k}^\prime\beta}),\label{eq:asymmetric-scattering-rate}
\end{align}
where $c(\xi_{\bm{k}\alpha})=\left(-\frac{J}{2N}\right)^2 \left[1-4\left(-\frac{J}{2N}\right)g(\xi_{\bm{k}\alpha})\right]\chi^z_{i j}$ with $g(\xi)=\Re \sum_{\bm{p}} \frac{f_{\bm{p}}-1/2}{\xi_{\bm{p}}-\xi}$, $\Delta\bm{k}=\bm{k}-\bm{k}^\prime$, $\bm{R}_{ij}=\bm{R}_i-\bm{R}_j$, and
$\bm{\chi}_{ij}=\langle\bm{S}_i\times\bm{S}_j\rangle$ ($i\ne j$) is the thermal average of the vector spin chirality (See Supplemental Material~\cite{suppl} for details).
Compared to a previous work on classical spins~\cite{Ishizuka2020a}, which analyzed the nonreciprocal current for classical spins, the higher-order terms in $J$ introduce corrections to the coefficient $c(\xi_{\bm{k}\alpha})$ through terms involving $g(\xi_{\bm{k}\alpha})$.
However, we find that the momentum and spin dependence expressed as $i\sigma^y_{\alpha\beta}\sin(\Delta\bm{k}\cdot\bm{R}_{ij})$ remains unchanged.
The $g(\xi)$ term, identical to the $\log(T)$ term in the Kondo effect~\cite{Kondo1964a}, arises from the non-commutativity of spin-flip processes, such as $S^{\pm}S^z$ and $S^z S^{\pm}$.
Its physical origin is the interference between the first- and second-order perturbations, which represents the interference between two distinct pathways: one where an electron is scattered once by a spin, and another where it is scattered twice by a spin.
It is precisely this double scattering by the same spin that gives rise to the Kondo effect.
Therefore, the $\log(T)$-term is a manifestation of the Kondo effect in the nonreciprocal response.

In contrast, such a quantum correction is absent in the anomalous Hall effect, at least to the third order in $J$~\cite{suppl}.
In addition, Onsager's reciprocal relation precludes any contribution from the fourth order.
Therefore, the quantum correction to the anomalous Hall effect is expected to be significantly weaker compared to the nonreciprocal response.

{\it Semiclassical Boltzmann theory.---}To gain further insight into the nonreciprocal response, we compute the nonreciprocal conductivity using Eq.~\eqref{eq:asymmetric-scattering-rate}.
Here, we phenomenologically interpret Eq.~\eqref{eq:asymmetric-scattering-rate} as the scattering probability of a single electron and use the semiclassical Boltzmann equation as in Ref.~\cite{Ishizuka2020a}.

For simplicity, we focus on the scattering by two neighboring spins aligned along the $z$-axis.
The electric field $\bm{E}$ and magnetic field are also applied along the $z$-axis.
The steady-state Boltzmann equation reads
\begin{equation}
\begin{split}
    -e\bm{E}\cdot&\frac{\partial f_{\bm{k}\alpha}}{\partial \bm{k}}
    =
    -\frac{f_{\bm{k}\alpha}-f^0_{\bm{k}\alpha}}{\tau}\\
    &+\sum_{\bm{k}^\prime\beta}[W^-_{\bm{k}^\prime\beta\to\bm{k}\alpha}f_{\bm{k}^\prime\beta}-W^-_{\bm{k}\alpha\to\bm{k}^\prime\beta}f_{\bm{k}\alpha}]
    .\label{eq:boltzmann-equation}
\end{split}
\end{equation}
The left-hand side represents the driving term from the external electric field $\bm{E}$, while the right-hand side is the collision integral, where the symmetric part of the scattering is treated within the relaxation-time approximation and the asymmetric part $W^-$ is treated explicitly.
To evaluate the second-order nonlinear response, we expand the nonequilibrium distribution function $f_{\bm{k}\alpha}$ in powers of both the electric field $\bm{E}$ and the asymmetric scattering rate $W^{-}$~\cite{Ishizuka2020a}.
The second-order nonlinear current we compute arises from the leading-order term in this expansion, which is second order in $\bm{E}$ and first order in $W^-$.
The resulting second-order nonlinear conductivity is 
\begin{equation}
\begin{split}
    &\frac{\sigma^{(2)}_{z z z}}{\sigma^2}
    \approx
    2\frac{\tau a^2}{e}\left(-\frac{J}{2\mu}\right)^2\frac{M\chi^z}{\mu}\frac{18\pi}{5} (k_F a)^2\\
    &\left(
        1
        +\frac{5}{2}J\rho
        \left(
            \log\left(\frac{T}{\mu}\right)
            -\frac{4}{5}\sqrt{\frac{\Lambda}{\mu}}
                +\frac{7}{5}-\gamma+\log\left(\frac{\pi}{8}\right)
        \right)
    \right),
    \label{eq:boltzmann-second-order-conductivity}
\end{split}
\end{equation}
We take the $k_F a \ll 1$ limit in the calculation.

The results of the semiclassical Boltzmann theory in this limit can be compared with those obtained from the Green's function method (Eq.~\eqref{eq:sigmazzz-green}). 
While the functional forms are almost identical to the Green's function result in the small-$k_F a$ limit, the numerical coefficients differ slightly.
By analogy with linear response theory---where the Boltzmann approach is equivalent to a Green's function calculation including ladder-type vertex corrections---this discrepancy is presumably related to the neglected vertex corrections in the Green's function method.

{\it Temperature Dependence.---}In addition to the cases without long-range magnetic order, such as partially-ordered states~\cite{Yokouchi2017a}, the Kondo-like enhancement may be observable in a magnet with an ordered phase.
To gain a better understanding of the nonreciprocal response in such systems, we next investigate the temperature dependence of the vector spin chirality.
The overall temperature dependence of the nonreciprocal conductivity involves not only a logarithmic $T$-dependence but also contributions from the thermal evolution of the magnetization $m$, the spin chirality $\chi^z$, and the relaxation time $\tau$.
We consider a ferromagnetic Heisenberg model with the Dzyaloshinskii--Moriya interaction and an external magnetic field on a three-dimensional simple cubic lattice~\cite{Buhrandt2013a,Yi2009a}, which is characterized by the exchange interaction $j$, Dzyaloshinskii--Moriya interaction $d$, and an external magnetic field $h$.
We focus on the paramagnetic phase, 
and study the spin correlations of an $S=1/2$ model using the high-temperature expansion.
In addition, we also employ Onsager's reaction field theory~\cite{Onsager1936,Brout1967,Logan1995,Wysin2000,Matsuura2003}
for classical spins, which should be a valid approximation for
large-$S$ quantum spins. 
The technical details and derivations are summarized in Supplemental Material~\cite{suppl}.

The resulting product $m\chi^z$, combined with our result for the conductivity coefficient $1+2J\rho\log\left(\frac{T}{\Lambda}\right)$, is shown in Fig.~\ref{fig:kondo_Tdep}.
As shown in the figure,  $\sigma^{(2)}$ increases as the temperature decreases, reflecting the development of spin correlations (represented by $m\chi^z$) in the paramagnetic phase above the magnetic transition temperature.
On top of the enhancement by the spin correlation, corrections such as the logarithmic temperature dependence are expected to further boost the nonreciprocal conductivity, which reaches about 180\% at $2T_K$, and 160\% at $4T_K$, where $\Lambda=4$, $J\rho=-0.15$ and $T_K=\Lambda\exp[1/J\rho]$.
This suggests a scenario where both quantum and thermal fluctuations cooperatively enhance the nonreciprocal response.

{\it Discussion.---}In this work, we theoretically investigated the effect of quantum fluctuations on the electrical magnetochiral effect arising from magnetic scattering, employing both the Green's function method and semiclassical Boltzmann theory.
The results reveal that a logarithmic enhancement of the nonreciprocal response occurs due to quantum fluctuations through a mechanism similar to the Kondo effect.
This temperature dependence is distinct from those arising from the increase of relaxation time $\tau$ and spin chirality $\bm{\chi}$ at low temperatures.
The results show a nontrivial consequence of quantum fluctuations in the nonlinear response and its relation to the Kondo effect.
Additionally, we show that the sign of the nonreciprocal conductivity is sensitive to band filling, similar to the anomalous Hall effect~\cite{Terasawa2025a}.
Our results indicate that a nontrivial effect unique to quantum fluctuations exists in the transport phenomena related to spin chirality.

Regarding experiments, recent studies of a partially-ordered state in MnSi under hydrostatic pressure~\cite{Pfleiderer2007a} have observed a topological Hall effect~\cite{Ritz2013a} and an enhanced nonreciprocal response~\cite{Yokouchi2017a}.
Quantum spin fluctuations have been discussed as a potential cause of these phenomena.
The enhanced nonreciprocal response aligns with our theoretical results.
Another promising platform is a nonmagnetic and noncentrosymmetric conducting material, such as RhGe, RhSi, CoGe, CoSi, and FeSi, doped with magnetic impurities, such as Mn~\cite{Sidorov2018a}.
Such a system serves as a noncentrosymmetric version of the Kondo problem; if two impurity spins are sufficiently close, the Dzyaloshinskii--Moriya interaction between them can induce chiral spin correlations. 
Confirming the logarithmic temperature dependence would establish the quantum correction and demonstrate that the nonreciprocal response serves as a probe for subtle quantum many-body correlations.

\acknowledgements

We are grateful to L. Balents for fruitful discussions.
This work was supported by JSPS KAKENHI (Grant Numbers JP23K03275, JP25H00841) and JST PRESTO (Grant No. JPMJPR2452).
%
%
\bibliography{ref} 

\end{document}